\documentclass[
twocolumn,
preprintnumbers,
nofootinbib,
showpacs,
prd,aps
]{revtex4}

\usepackage{epsf}

\begin{document}


\renewcommand{\topfraction}{0.8}

\preprint{TU-760}

\title{
Right-Handed Sneutrino as Cold Dark Matter
}

\author{
Takehiko Asaka, Koji Ishiwata and Takeo Moroi
}

\affiliation{
Department of Physics, Tohoku University, 
Sendai 980-8578, Japan
}

\date{December 9, 2005}

\begin{abstract}

    We consider supersymmetric models with right-handed neutrinos
    where neutrino masses are {\it purely Dirac-type}.  In this model,
    right-handed sneutrino can be the lightest supersymmetric particle
    and can be a viable candidate of cold dark matter of the universe.
    Right-handed sneutrinos are never thermalized in the early
    universe because of weakness of Yukawa interaction, but are
    effectively produced by decays of various superparticles.  We show
    that the present mass density of right-handed sneutrino can be
    consistent with the observed dark matter density.

\end{abstract}

\maketitle

In recent years, various experiments have confirmed the phenomenon of
neutrino oscillation.  (See, for example,
\cite{SK_S,SK_A,K2K,SNO,KamLAND}.)  Those results strongly suggest
very small but non-vanishing neutrino masses.  This fact raises
serious problems because the non-vanishing neutrino mass is not
allowed in the standard model of particle physics and also because
suggested values of neutrino masses are extremely small.  The easiest
way of generating neutrino masses is to introduce right-handed
neutrinos; with this extension, Yukawa couplings of neutrinos may
exist.  Consequently, neutrinos can acquire masses after electroweak
symmetry breaking.

Even with right-handed neutrinos, there are two different classes of
scenarios for generating neutrino masses.  Probably, more popular one
is with Majorana masses for right-handed neutrinos, i.e., so-called
``seesaw'' scenario \cite{seesaw}.  In this scenario, smallness of the
neutrino masses is explained by the Majorana masses of right-handed
neutrinos which are much larger than the electroweak scale.

Small neutrino masses can be, however, realized without seesaw
mechanism.  With vanishing Majorana masses of the right handed
neutrinos, which may be the consequence of exact lepton-number
symmetry, neutrino masses become Dirac-type.  As we will see, in this
case, Yukawa coupling constants for neutrinos are, roughly speaking,
$O(10^{-13})$ or smaller to make the neutrino masses to be consistent
with the results of neutrino-oscillation experiments.  One might think
that such small Yukawa coupling constants are unnatural.  It is,
however, natural in 't Hooft's sense \cite{thooft} since some symmetry
(i.e., chiral symmetry in neutrino sector) is restored in the limit of
vanishing neutrino Yukawa coupling constants.  In this model, however,
right-handed neutrinos are mostly irrelevant for collider experiments
and cosmology since their interaction is extremely weak.

In supersymmetric models, which is strongly motivated as a solution to
various problems in the standard model of particle physics, like
hierarchy and naturalness problems, situation changes.  In particular,
superpartners of right-handed neutrinos may play an important role in
cosmology.  When small neutrino masses are purely Dirac-type, masses
of right-handed sneutrinos are dominantly from effects of
supersymmetry (SUSY) breaking.  Then, one should note that the
lightest superparticle (LSP) may be the (lightest) right-handed
neutrino $\tilde{\nu}_R$.  Importantly, since the LSP $\tilde{\nu}_R$
becomes stable by the $R$-parity conservation and also is very weakly
interacting, it can be a viable candidate of cold dark matter (CDM)
provided that its relic density is the right amount.

In this letter, we consider the minimal supersymmetric standard model
(MSSM) with three generations of right-handed (s)neutrinos where small
neutrino masses are purely Dirac-type.  In particular, we study the
case where the LSP is the lightest right-handed sneutrino and see if
the relic density of $\tilde{\nu}_R$ can become consistent with the
present CDM density.  Since interaction of right-handed sneutrino is
very weak, it is not thermalized in thermal bath.  Even in this case,
some of decay (and scattering) processes produce $\tilde{\nu}_R$.  As
we will see, in some parameter region, the density parameter of
right-handed sneutrino $\Omega_{\tilde{\nu}_R}$ can be $O(0.1)$, which
is consistent with the CDM density suggested by the WMAP \cite{WMAP}:
\begin{eqnarray}
    \Omega_{\rm c}h_{100}^2=0.1126^{+0.0161}_{-0.0181},
\end{eqnarray}
where $h_{100}$ is the Hubble constant in units of $100\ {\rm
km/sec/Mpc}$.

Let us first introduce interaction and mass terms in Lagrangian.  The
important part of superpotential is
\begin{eqnarray}
    W = y_\nu \hat{H}_u \hat{L} \hat{\nu}_R^c 
    + \mu_H \hat{H}_u \hat{H}_d,
\end{eqnarray}
where $\hat{H}_u=(\hat{H}_u^+, \hat{H}_u^0)$ and
$\hat{H}_d=(\hat{H}_d^0, \hat{H}_d^-)$ are up- and down-type Higgses,
and $\hat{L}=(\hat{\nu}_L, \hat{l}_L^-)$ the left-handed lepton.  (In
this letter, ``hat'' is for superfields while ``tilde'' is for
superparticles with odd $R$-parity.)  Here, we omit flavor indices for
simplicity.  With this superpotential, neutrino mass is generated
after electroweak symmetry breaking: $m_{\nu}=y_\nu \langle H_u^0
\rangle = y_\nu v\sin\beta$, where $v\simeq 174\ {\rm GeV}$ and $\tan
\beta = \langle H_u^0 \rangle/ \langle H_d^0 \rangle$.  Thus, the
Yukawa coupling constant $y_\nu$ is determined once the neutrino mass
is fixed:
\begin{eqnarray}
    y_\nu \sin\beta = 3.0 \times 10^{-13} \times
    \left( \frac{m_{\nu}^2}{2.8 \times 10^{-3}\ {\rm eV^2}} 
    \right)^{1/2}.
\end{eqnarray}
For simplicity, we consider the case where $\tan\beta$ is relatively
large.  In this case $H_u$ behaves like the standard-model Higgs;
$H_u\simeq H_{\rm SM}$.  The lightest Higgs boson $h$ is contained in
$H_{\rm SM}$, and we take its mass to be $m_h=115\ {\rm GeV}$ in our
study.  On the contrary, $H_d$ plays the role of heavy Higgs doublet.
In addition, we approximate mass eigenstates of the superparticles by
left- and right-handed sleptons, gauginos, and Higgsinos.  Effects of
left-right mixing are taken into account by mass-insertion method.
The mass terms for superpaticles are given by
\begin{eqnarray}
    {\cal L}_{\rm mass} &=& 
    - \frac{1}{2}
    \left[ m_{\tilde{B}} \tilde{B}\tilde{B}
        + m_{\tilde{W}} \tilde{W}\tilde{W}
        + \mu_H \tilde{H}_u \tilde{H}_d
        + {\rm h.c.} \right]
    \nonumber \\ &&
    - m_{\tilde{\nu}_R}^2 \tilde{\nu}_R^* \tilde{\nu}_R
    - m_{\tilde{l}_L}^2 \tilde{l}_L^* \tilde{l}_L
    - m_{\tilde{\nu}_L}^2 \tilde{\nu}_L^* \tilde{\nu}_L,
\end{eqnarray}
where $\tilde{B}$ and $\tilde{W}$ are gauginos for $U(1)_Y$ and
$SU(2)_L$ gauge groups, respectively.  Notice that all the terms
except for Higgsino mass term are from SUSY breaking
and that $m_{\tilde{l}_L}^2$ and $m_{\tilde{\nu}_L}^2$ are identical
up to $D$-term contribution.  Furthermore, we introduce soft-SUSY
breaking tri-linear coupling which will become important in the
following discussion
\begin{eqnarray}
    {\cal L}_{A} = A_\nu H_u \tilde{L} \tilde{\nu}_R^c + {\rm h.c.}
\end{eqnarray}
For later convenience, we parameterize
\begin{eqnarray}
    m_{\tilde{l}_L, \tilde{\nu}_L}^2 = m_{\tilde{L}}^2
    + \Delta^{(D)}_{\tilde{l}_L, \tilde{\nu}_L},~~~
    A_\nu &=& a_\nu y_\nu m_{\tilde{L}},
\end{eqnarray}
where $\Delta^{(D)}_{\tilde{l}_L, \tilde{\nu}_L}$ represents the
$D$-term contribution.  In addition, we adopt the GUT relation among
the gaugino masses: $m_{\tilde{B}}\simeq 0.5m_{\tilde{W}}$.

Now let us discuss the production of the LSP right-handed sneutrino
$\tilde {\nu}_R$ via various decay processes.  Since we found that the
production via scatterings is subdominant and can be neglected, we
consider here only the decay processes $x \rightarrow \tilde
{\nu}_Ry$.  We assume that particles $x$ and $y$ are in chemical
equilibrium.  Their distribution functions at the temperature $T$,
then, are given by $f(E)=(e^{E/T}\pm 1)^{-1}$, where the positive and
negative signs are for fermions and bosons, respectively.

The evolution of the number density of $\tilde {\nu}_R$ 
is governed by the Boltzmann equation
\begin{eqnarray}
    \dot{n}_{\tilde{\nu}_R} + 3 H n_{\tilde{\nu}_R}
    = C_{\rm decay},
    \label{BoltzmannEq}
\end{eqnarray}
where ``dot'' represents time-derivative and $H$ is the Hubble parameter.
The source term from the decay process is written in the form
\begin{eqnarray}
  C_{\rm decay} = 
  \sum_{x,y} \int 
    \frac{d^3 k_x}{(2\pi)^3}
    \gamma_x (2 s_x + 1)
    \Gamma_{x\rightarrow \tilde{\nu}_R y}
    f_x
    \langle 1 \pm f_y \rangle_{k_x},
\end{eqnarray}
where $\gamma_x=m_x/\sqrt{k_x^2+m_x^2}$ is the Lorentz factor and
$(2s_x+1)$ is the spin multiplicity of $x$.  Furthermore, $\langle 1
\pm f_y \rangle_{k_x}$ is the averaged final-state multiplicity factor
for fixed value of initial-state momentum.  (Here, the positive and
negative signs are for bosons and fermions, respectively.)

Decay rates for processes which become important for the calculation
of $\Omega_{\nu_R}$ are given by
\begin{eqnarray}
    && 
    \Gamma_{\tilde{H}^0 \rightarrow \tilde{\nu}_R \bar{\nu}_L} 
    = 
    \Gamma_{\tilde{H}^+ \rightarrow \tilde{\nu}_R l^+_L} 
    =
    \frac{\beta_{\rm f}^2 y_\nu^2}{32\pi} \mu_H ,
    \label{Gamma_hgg}
    \\ &&
    \Gamma_{\tilde{\nu}_L\rightarrow \tilde{\nu}_R h} 
    =
    \frac{\beta_{\rm f}}{32\pi} \frac{A_\nu^2}{m_{\tilde{\nu}_L}},
    \\ &&
    \Gamma_{\tilde{\nu}_L\rightarrow \tilde{\nu}_R Z} 
    =
    \frac{\beta_{\rm f}^3}{32\pi} 
    \left[\frac{m_{\tilde{\nu}_L}^2}
        {m_{\tilde{\nu}_L}^2-m_{\tilde{\nu}_R}^2}
    \right]^2 
    \frac{A_\nu^2}{m_{\tilde{\nu}_L}},
    \\ &&
    \Gamma_{\tilde{l}_L\rightarrow \tilde{\nu}_R W^-} 
    =
    \frac{\beta_{\rm f}^3}{16\pi} 
    \left[\frac{m_{\tilde{l}_L}^2}
        {m_{\tilde{\nu}_L}^2-m_{\tilde{\nu}_R}^2}
    \right]^2 
    \frac{A_\nu^2}{m_{\tilde{l}_L}},
    \\ &&
    \Gamma_{\tilde{B}\rightarrow \tilde{\nu}_R \bar{\nu}_L} 
    =
    \frac{\beta_{\rm f}^2 g_1^2}{64\pi}
    \left[\frac{A_\nu v_T}
        {m_{\tilde{\nu}_L}^2-m_{\tilde{\nu}_R}^2}
    \right]^2
    m_{\tilde{B}},
    \label{Gamma_bino}
    \\ &&
    \Gamma_{\tilde{W}^0\rightarrow \tilde{\nu}_R \bar{\nu}_L}
    =
    \frac{\beta_{\rm f}^2 g_2^2}{64\pi}
    \left[\frac{A_\nu v_T}
        {m_{\tilde{\nu}_L}^2-m_{\tilde{\nu}_R}^2}
    \right]^2
    m_{\tilde{W}},
    \\ &&
    \Gamma_{\tilde{W}^+ \rightarrow \tilde{\nu}_R l^+_L} 
    =
    \frac{\beta_{\rm f}^2 g_2^2}{32\pi}
    \left[\frac{A_\nu v_T}
        {m_{\tilde{\nu}_L}^2-m_{\tilde{\nu}_R}^2}
    \right]^2
    m_{\tilde{W}},
    \label{Gamma_wino+-}
\end{eqnarray}
where we have neglected lepton masses.  Here, $g_1$ and $g_2$ are
gauge coupling constants for the $U(1)_Y$ and $SU(2)_L$ gauge groups,
respectively, and, for the process $x\rightarrow \tilde{\nu}_Ry$,
$\beta_{\rm f}$ is given by
\begin{eqnarray}
    \beta_{\rm f}^2 = \frac{1}{m_x^4}
    [ m_x^4 - 2 (m_{\tilde{\nu}_R}^2 + m_y^2) m_x^2
    + (m_{\tilde{\nu}_R}^2 - m_y^2)^2 ],
\end{eqnarray}
with $m_x$ and $m_y$ being the masses of the particles $x$ and $y$,
respectively.  In addition, $v_T$ is temperature-dependent Higgs VEV.
We approximate the Higgs potential in thermal bath as
\cite{Dine:1992wr}
\begin{eqnarray}
    V_T &\simeq& \frac{m_h^2}{4v^2} ( |H_{\rm SM}|^2 - v^2 )^2
    \nonumber \\ &&
    + \frac{1}{8v^2} (2m_W^2 + m_Z^2 + 2m_t^2) T^2 |H_{\rm SM}|^2,
\end{eqnarray}
where $T$ is cosmic temperature.  Then we minimize $V_T$ to obtain
$v_T\simeq \langle H_u\rangle_T$.

In solving Eq.~(\ref{BoltzmannEq}) it is useful to define the yield
variable
\begin{eqnarray}
  Y_{\tilde{\nu}_R} \equiv \frac{n_{\tilde{\nu}_R}}{s},
\end{eqnarray}
where $s\equiv\frac{2\pi^2}{45}g_*T^3$ is the total entropy density of
the universe. (In this analysis, we take the effective number of
massless degrees of freedom as $g_\ast = 106.75$, since the production
of $\tilde{\nu}_R$ becomes effective at the temperature lower than
most of superparticle masses.)

Using the relation $\dot{T}=-HT$, $Y_{\tilde{\nu}_R}$ is given by
\begin{eqnarray}
    Y_{\tilde{\nu}_R} (T) = 
    \int_{T}^{T_{\rm max}}
    \frac{C_{\rm decay}}{sHT}  dT.
    \label{Y_nuR}
\end{eqnarray}
Once the yield variable is obtained, we can also calculate the mass
density of sneutrino using the relation
$\rho_{\tilde{\nu}_R}/s=m_{\tilde{\nu}_R}Y_{\tilde{\nu}_R}$.
Comparing with the present value of the critical density $\rho_{\rm
crit}$, which is given by $[\rho_{\rm crit}/s]_{\rm now}\simeq3.6\ 
h_{100}^2\times 10^{-9}\ {\rm GeV}$,
we obtain density parameter of right-handed sneutrino
$\Omega_{\tilde{\nu}_R}\equiv (\rho_{\tilde{\nu}_R}
+\rho_{\tilde{\nu}^*_R})/\rho_{\rm crit}$.

Importantly the present value of $Y_{\tilde{\nu}_R}$ is insensitive to
the maximal temperature $T_{\rm max}$ of the universe.  To see this,
it is instructive to roughly estimate $Y_{\tilde{\nu}_R}$.  Neglecting
the Lorentz factor, let us approximate the decay term as $C_{\rm
decay}\sim \frac{N_{\rm mode}}{16\pi}y_\nu^2m_{\rm SUSY}n_{\rm SUSY}$,
where $m_{\rm SUSY}$ and $n_{\rm SUSY}$ are typical mass scale and
number density of (parent) superparticles, respectively, and $N_{\rm
mode}$ is the number of possible decay mode.  The number density
$n_{\rm SUSY}$ is $\sim T^3$ when the temperature is higher than
$m_{\rm SUSY}$, and is exponentially suppressed when $T\ll m_{\rm
SUSY}$.  The integration in Eq.\ (\ref{Y_nuR}) is then dominated at
the temperature of $T\sim m_{\rm SUSY}$.  Consequently, we obtain
$Y_{\tilde{\nu}_R}(T\ll m_{\rm SUSY})\sim \frac{N_{\rm mode}y_\nu^2
M_*}{16\pi g_*^{3/2}m_{\rm SUSY}}$, with $M_*\simeq 2.4\times 10^{18}\ 
{\rm GeV}$ being the reduced Planck scale.  As one can see, the
present $Y_{\tilde{\nu}_R}$ is insensitive to thermal history for
$T\gg m_{\rm SUSY}$.  Thus, $\Omega_{\tilde{\nu}_R}$ does not depend
on, for example, reheating temperature after inflation.  We can also
estimate the density parameter; using the above estimation of
$Y_{\tilde{\nu}_R}$, we obtain $\Omega_{\tilde{\nu}_R}h_{100}^2\sim
O(10^{-3}) \times N_{\rm mode} (\frac{y_\nu}{3\times 10^{-13}})^2
(\frac{m_{\tilde{\nu}_R}}{m_{\rm SUSY}})$.  With this naive
estimation, $\Omega_{\tilde{\nu}_R}$ becomes smaller than the
currently observed CDM density.  In some case, however,
$\Omega_{\tilde{\nu}_R}$ becomes much larger, as we see below.

Now, we are at the position to quantitatively estimate
$\Omega_{\tilde{\nu}_R}$.  We have numerically evaluated the yield
variable using Eq.\ (\ref{Y_nuR}) for several choices of parameters.
The relic density of right-handed sneutrino strongly depends on the
neutrino Yukawa coupling constant which is related to the neutrino
mass.  Neutrino-oscillation experiments determine only the
mass-squared differences of neutrinos; here we adopt
\cite{K2K,KamLAND}
\begin{eqnarray}
    \left[ \Delta m_{\nu}^2 \right]_{\rm atom}
    &=& 2.8 \times 10^{-3}\ {\rm eV^2},
    \label{dm_atom}
    \\
    \left[ \Delta m_{\nu}^2 \right]_{\rm solar}
    &=& 7.9 \times 10^{-5}\ {\rm eV^2}.
    \label{dm_solar}
\end{eqnarray}

First, let us consider the case where the masses of the neutrinos are
hierarchical.  In this case, $\left[ \Delta m_{\nu}^2 \right]_{\rm
atom}$ almost corresponds to the mass-squared of the heaviest neutrino
which we call third generation neutrino.  In this case, the largest
Yukawa coupling constant is given by $y_\nu^{(3)}\simeq 3\times
10^{-13}$, and other Yukawa coupling constants are much smaller.
(Here and hereafter, the superscript ``$(i)$'' indicates that
$y_\nu^{(i)}$ is for neutrino in $i$-th generation.)  As a result,
production of $\tilde{\nu}_R$ is dominated by processes where the
third-generation (s)neutrino is related.

As show in in Eqs.\ (\ref{Gamma_hgg}) $-$ (\ref{Gamma_wino+-}), there
exist various decay processes which produce $\tilde{\nu}_R$.  Among
them, Higgsino decay process dominates the $\tilde{\nu}_R$ production
when the effects of the tri-linear scalar couplings are negligible.
In this case, however, $\Omega_{\tilde{\nu}_R}$ becomes too small to
be consistent with observation of the CDM relic density.  Indeed, when
$m_{\tilde{\nu}_R}=100\ {\rm GeV}$, for example, we found
$\Omega_{\tilde{\nu}_R}= 0.004 - 0.001$ for $\mu_H=200\ {\rm GeV} - 1\ 
{\rm TeV}$, which is much smaller than the present CDM density.  

If some of other decay processes are effective,
$\Omega_{\tilde{\nu}_R}$ can become significantly larger.  In
particular, when the tri-linear coupling constant $A_\nu$ is
non-vanishing, decays of various superparticles produces
$\tilde{\nu}_R$ sufficiently.

\begin{figure}[t]
    \begin{center}
        \epsfxsize=0.45\textwidth\epsfbox{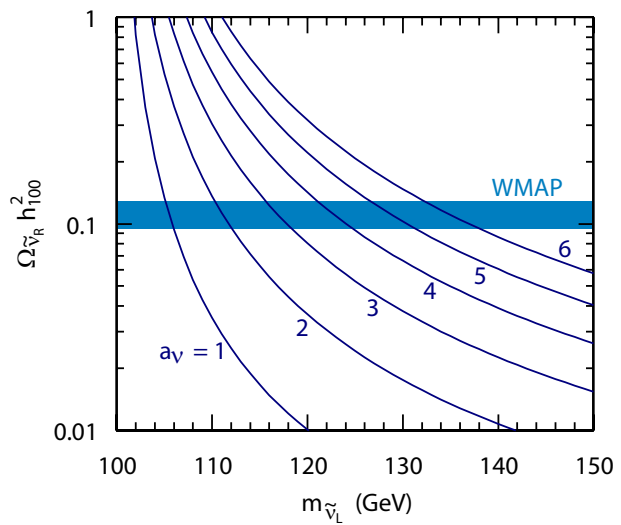}
        \caption{$\Omega_{\tilde{\nu}_R}h_{100}^2$ as a function of 
        $m_{\tilde{\nu}_L}$ for $a_\nu=1-6$.  Here, we take
        $m_{\tilde{\nu}_R}=100\ {\rm GeV}$, $m_{\tilde{W}}=300\ {\rm
        GeV}$, and $\mu_H=150\ {\rm GeV}$.  The shaded band
        corresponds to the CDM density suggested by the WMAP.}
        \label{fig:Omg_winodec}
    \end{center}
    \vspace{-0.5cm}
\end{figure}

One possible enhancement can be due to the process
$\tilde{W}\rightarrow \tilde{\nu}_R+\cdots$; if the mass of
left-handed sneutrino $\tilde{\nu}_L$ is close to $m_{\tilde{\nu}_R}$,
production of $\tilde{\nu}_R$ is enhanced because the left-right
mixing of sneutrinos becomes larger as
$|m_{\tilde{\nu}_L}^2-m_{\tilde{\nu}_R}^2|$ becomes smaller.  (See
Eqs.\ (\ref{Gamma_bino}) $-$ (\ref{Gamma_wino+-}).) In fact,
$\Omega_{\tilde{\nu}_R}$ can be of $O(0.1)$ with a mild degeneracy of
$\tilde{\nu}_R$ and $\tilde{\nu}_L$.  To see this, in Fig.\ 
\ref{fig:Omg_winodec}, we show $\Omega_{\tilde{\nu}_R}h_{100}^2$ as a
function of $m_{\tilde{\nu}_L}$.  We can see that $10-20\ \%$
degeneracy is enough to realize $\tilde{\nu}_R$-CDM even if $a_\nu\leq
3$.  Here, we take $m_{\tilde{W}}=300\ {\rm GeV}$.  If we increase the
Wino mass, $\Omega_{\tilde{\nu}_R}$ decreases since the
temperature-dependent VEV $v_T$ becomes suppressed at high
temperature.  We also note that the process
$\tilde{H}^+\rightarrow\tilde{\nu}_R\tau_R^+$ gives additional
contribution to $\Omega_{\tilde{\nu}_R}$, which may become sizable
when $\tan\beta$ is large.  We found that $\Omega_{\tilde{\nu}_R}$ can
be $\sim 40 \%$ larger for $\tan\beta =55$.

Even without mass degeneracy between $\tilde{\nu}_R$ and
$\tilde{\nu}_L$, $\tilde{\nu}_R$-CDM is realized if $a_\nu\gg 1$.
Although such a very large value of $a_\nu$ may not be realized in
simple supergravity models, it is phenomenologically viable.

Next, we turn to consider the case of degenerate neutrino masses,
where there is another possibility of enhancing
$\Omega_{\tilde{\nu}_R}$.  In this case, the relation
$y_\nu^{(1)}\simeq y_\nu^{(2)}\simeq y_\nu^{(3)}$ holds and all three
generations of right-handed sneutrinos may be effectively produced.
The point is that the neutrino Yukawa coupling constants can be much
larger than $\sim 3\times 10^{-13}$, and hence
$\Omega_{\tilde{\nu}_R}$ can be more enhanced than the hierarchical
case. $\tilde{\nu}_R$-CDM can be realized if the neutrino masses are
$O(0.1\ {\rm eV})$ even if there is no other enhancement.

So far, we have only discussed the production of $\tilde{\nu}_R$ by
the decay of superparticles {\it in chemical equilibrium}.  In
general, $\tilde{\nu}_R$ can be also produced after superparticles
freeze out.  The lightest superparticle in the MSSM sector, which we
call MSSM-LSP, is very long-lived because of the weakness of the
neutrino Yukawa interaction.  Thus, decay of the MSSM-LSP occurs after
its number is frozen, which gives another source of $\tilde{\nu}_R$.
The total density parameter is then given by
$\Omega_{\tilde{\nu}_R}^{\rm (total)}=
\Omega_{\tilde{\nu}_R}+\frac{m_{\tilde{\nu}_R}}{m_{\rm
MSSM}}\Omega_{\rm MSSM}$, where $m_{\rm MSSM}$ is the MSSM-LSP mass.
Here, the first term is evaluated with Eq.\ (\ref{Y_nuR}) while the
second one is from the decay after the freeze out; $\Omega_{\rm MSSM}$
is the expected density parameter of the MSSM-LSP for the case where
the MSSM-LSP is stable.  Notice that $\Omega_{\rm MSSM}$ is
model-dependent and may be smaller than the WMAP value; if
$\Omega_{\rm MSSM}$ is large, the parameter space to realize
$\tilde{\nu}_R$-CDM is enlarged.

Here we discuss the effects of the MSSM-LSP decay on big-bang
nucleosynthesis (BBN).  If the MSSM-LSP decays after BBN starts, it
may cause hadro- and photo-dissociation of light elements, which may
spoil the success of standard BBN scenario.  In order to realize
$\tilde{\nu}_R$-CDM with enhanced left-right mixing, $\tilde{\nu}_L$
is required to be relatively light and then it may be the MSSM-LSP.
If so, we found that, for $m_{\tilde{\nu}_R}=100\ {\rm GeV}$, its
lifetime becomes so long that light-element abundances become
inconsistent with observations due to hadro-dissociation processes
\cite{KawKohMor} unless $m_{\tilde{\nu}_L}\gtrsim 140\ {\rm GeV}$
(with $a_\nu \gtrsim 6$). On the other hand, very degenerate
sneutrinos ($m_{\tilde{\nu}_L}-m_{\tilde{\nu}_R}\lesssim$ a few GeV)
is another possibility to avoid the difficulty by suppressing the
energy release from $\tilde{\nu}_L$ decay.  Otherwise, the BBN
constraints can be also avoided if the MSSM-LSP is the lightest
neutralino, or if $R$-parity in the MSSM sector is (very weakly)
broken.

Finally, we comment on the decay of heavier right-handed sneutrinos
(denoted by $\tilde{\nu}'_R$).  When the mass difference between
$\tilde{\nu}'_R$ and $\tilde{\nu}_R$ is large enough, $\tilde{\nu}'_R$
may decay, for example, into $\tilde{H}_u L'$, and the produced
Higgsino decays subsequently as $\tilde{H}_u\rightarrow\tilde{\nu}_R
L$.  In this case, lifetime of the long-lived particles can be $\sim
1\ {\rm sec}$ or shorter; such decay processes are cosmologically
safe~\cite{KawKohMor}.  On the contrary, if $\tilde{\nu}'_R$ decays
only through the processes of
$\tilde{\nu}'_R\rightarrow\tilde{\nu}_R+\cdots$ due to kinematical
reason, the lifetime becomes much longer than the present age of the
universe.  For instance, when $m_{\tilde{\nu}_R}=100\ {\rm GeV}$,
$m_{\tilde{\nu}'_R}=150\ {\rm GeV}$, and $\mu_H=200$, $300$, and $400\ 
{\rm GeV}$, the lifetime of $\tilde{\nu}'_R$ is estimated as
$\tau_{\tilde{\nu}'_R}\simeq 5\times 10^{24}$, $5\times 10^{25}$, and
$2\times 10^{26}\ {\rm yr}$, respectively.  (Here, we approximated
that all the neutrino Yukawa coupling constants are $10^{-13}$, and
neglected tri-linear scalar couplings.)  Even in this case, some
(tiny) amount of heavier right-handed sneutrinos decay into charged
particles until today.  Assuming the density parameter of
$\tilde{\nu}'$ to be $O(0.01)$, flux of the (primary) charged
particles is estimated to be ${\cal F}\sim 10^{-16}\ {\rm
cm^{-2}sec^{-1}str^{-1}GeV^{-1}}\times
(\frac{\tau_{\tilde{\nu}'_R}}{10^{25}\ {\rm yr}})^{-1}$.  The emitted
charged particles scatter off the background radiation and produce
energetic photons.  However, we expect that such small flux is
consistent with the observations since the above flux is much smaller
than the observed $\gamma$-ray flux with the energy $10-100\ {\rm
GeV}$: ${\cal F}^{\rm (obs)}\sim 10^{-8}- 10^{-10}\ {\rm
cm^{-2}sec^{-1}str^{-1}GeV^{-1}}$ \cite{Sreekumar:1997un}.

In summary, in this letter, we have considered supersymmetric models
with the right-handed-sneutrino LSP, where masses of neutrinos are
purely Dirac-type.  We have seen that, in such models,
$\Omega_{\tilde{\nu}_R}$ can be of $O(0.1)$ and right-handed sneutrino
can be CDM.  In particular, if $\tilde{\nu}_R$-CDM is realized in the
parameter region where $\tilde{\nu}_R$ and $\tilde{\nu}_L$ are quite
degenerate, left-handed sleptons become fairly light.  Such light
sletons are interesting targets of the future collider experiments.
On the contrary, if $\tilde{\nu}_R$-CDM is realized with degenerate
neutrino masses, the (active) neutrino mass should become $\sim O(0.1\ 
{\rm eV})$.  Such a scenario may be tested by using matter power
spectrum from Ly-$\alpha$ forest data \cite{Seljak:2004xh}.

If $\tilde{\nu}_R$ is the LSP, there are much more to be discussed
from phenomenological and cosmological points of view.  Even though
the LSP is $\tilde{\nu}_R$, the MSSM-LSP looks like stable particle in
collider experiments.  However, in this case, the MSSM-LSP may be
charged or even colored.  In addition, cosmology may be also
significantly affected if the right-handed sneutrino is the LSP; one
of the examples has been discussed in this letter.  We also note that,
if the initial amplitude of $\tilde{\nu}_R$ is $\sim 10^9\ {\rm GeV}$,
coherent mode of the right-handed sneutrino may become the CDM.  We
believe that the scenario with right-handed-sneutrino LSP provides new
and rich phenomenology.

\noindent
{\it Acknowledgements:} This work was supported by the Grant-in-Aid
for Scientific Research from the Ministry of Education, Science,
Sports, and Culture of Japan, No.\ 16081202 (TA) and No.\ 15540247
(TM).


\vspace{-0.7cm}

\begin{thebibliography}{99}

\vspace{-1.7cm}

\bibitem{SK_S}
    S.~Fukuda {\it et al.},
    Phys.\ Lett.\ B {\bf 539}, 179 (2002).

\bibitem{SK_A}
    Y.~Ashie {\it et al.},
    Phys.\ Rev.\ D {\bf 71}, 112005 (2005).

\bibitem{K2K}
    E.~Aliu {\it et al.},
    Phys.\ Rev.\ Lett.\  {\bf 94}, 081802 (2005).

\bibitem{SNO}
    S.~N.~Ahmed {\it et al.},
    Phys.\ Rev.\ Lett.\  {\bf 92}, 181301 (2004).

\bibitem{KamLAND}
    T.~Araki {\it et al.},
    Phys.\ Rev.\ Lett.\  {\bf 94}, 081801 (2005).

\bibitem{seesaw}
    T.\ Yanagida, 
    in ``Proceedings of the Workshop on Unified Theory and Baryon
    Number of the Universe,'' 
    eds.\ O.\ Sawada and A.\ Sugamoto (KEK, 1979) p.95;
    M.\ Gell-Mann, P.\ Ramond and R.\ Slansky, 
    in ``Supergravity,''
    eds.\ P.\ van Niewwenhuizen and D.\ Freedman (North Holland,
    1979);
    S.~L.~Glashow,
    in ``Proceedings of the Carg\'ese Summer Institute on Quarks and
    Leptons,'' (Plenum, 1980) p707.

\bibitem{thooft}
    G.~'t Hooft,
    in ``Recent Developments in Gauge Theories,''
    (Plenum, 1980) p.135.

\bibitem{WMAP}
    D.~N.~Spergel {\it et al.},
    Astrophys.\ J.\ Suppl.\  {\bf 148}, 175 (2003).

\bibitem{Dine:1992wr}
    M.~Dine {\it et al.},
    Phys.\ Rev.\ D {\bf 46}, 550 (1992).

\bibitem{Seljak:2004xh}
    U.~Seljak {\it et al.},
    Phys.\ Rev.\ D {\bf 71}, 103515 (2005).

\bibitem{KawKohMor}
    M.~Kawasaki, K.~Kohri and T.~Moroi,
    Phys.\ Lett.\ B {\bf 625}, 7 (2005);
    Phys.\ Rev.\ D {\bf 71}, 083502 (2005).

\bibitem{Sreekumar:1997un}
    P.~Sreekumar {\it et al.},
    Astrophys.\ J.\  {\bf 494}, 523 (1998).

\end{thebibliography}
\end{document}